\documentclass[reprint,prb,twocolumn,showpacs,superscriptaddress,aps,longbibliography]{revtex4-1}
\usepackage[utf8]{inputenc}

\usepackage{graphicx}
\DeclareGraphicsExtensions{.png,.jpg,.eps,.pdf}

\usepackage{float}
\usepackage{xcolor}
\usepackage{amsmath}
\usepackage[printwatermark]{xwatermark}
\usepackage{float}
\usepackage[colorlinks=true,citecolor=blue]{hyperref}

\DeclareMathAlphabet\mathbfcal{OMS}{cmsy}{b}{n}

\begin{document}

\title{Ferroelectric valley valves with graphene/MoTe$_2$ van der Waals heterostructures}

\author{Adolfo O. Fumega}
\affiliation{Department of Applied Physics, Aalto University, 02150 Espoo, Finland}

\author{Jose L. Lado}
\affiliation{Department of Applied Physics, Aalto University, 02150 Espoo, Finland}

\begin{abstract}
Ferroelectric van der Waals heterostructures 
provide a natural platform to design a variety of electrically
controllable devices.
In this work, we demonstrate that AB bilayer graphene encapsulated in MoTe$_2$ acts as a valley valve that displays a switchable built-in
topological gap, leading to 
ferroelectrically driven
topological channels.
Using a combination of \emph{ab initio} calculations and low energy models, we show that 
the ferroelectric order of MoTe$_2$ allows the control of the gap opening in bilayer graphene and leads
to topological channels between different ferroelectric domains. Moreover, we analyze the effect that the moiré modulation between MoTe$_2$ and graphene layers has in the topological modes, 
demonstrating
that the edge states are robust against moiré modulations of the ferroelectrically-induced
electric potential. 
Our results put forward ferroelectric/graphene heterostructures as versatile platforms to engineer
switchable built-in topological channels
without requiring an external electric bias.
\end{abstract}

\maketitle

\section{Introduction}

Layered van der Waals compounds have become a fertile platform to engineer materials with
emergent functionalities,\cite{vdwHT2013} ranging from
fundamental quantum states\cite{PhysRevLett.124.106803,PhysRevResearch.2.023238,Xie2021} to
nanoelectronic devices
including field effect transistors\cite{Das2012,Wang2012,Lee2015} and valves.\cite{Chen2021valve,Lin2020,PhysRevLett.121.067701} 
The weak van der Waals bonding between layers 
allows to combine different monolayers with lattice constants, and to impose a rotation between
the materials. This flexibility permits to use 2D building blocks with different symmetry-breaking orders: ferromagnets,\cite{doi:10.1021/acs.nanolett.6b03052,Huang2017,Gong2017,Fei2018,doi:10.1021/acs.nanolett.9b00553} superconductors,\cite{NbSe22015} ferroelectrics\cite{Yuan2019} or multiferroics;\cite{Song2022, Fumega_2022} that can be wisely merged to create heterostructures with promising applications. Paradigmatic examples of such combinations of different electronic orders
are graphene/ferromagnet heterostructures
realizing van der Waals tunnel junctions and valley polarization,\cite{Klein2018,Jiang2019,ZHOU201825,QI2021540}
graphene/semiconductor heterostructures leading to strong spin-orbit coupling effects,\cite{PhysRevB.98.155309,PhysRevB.104.195156}
ferromagnet/superconductor heterostructures realizing topological superconductivity,\cite{Kezilebieke2020,Kezilebieke2022}
ferromagnet/semiconductor providing controllable excitons,\cite{PhysRevLett.124.197401,Lyons2020}
magnet/metal realizing heavy-fermion states,\cite{Vao2021,2022arXiv220700096W} and
ferroelectric/twisted bilayers realizing ferroelectrically
switchable superconductivity.\cite{2022arXiv220504458K}
In particular, 2D ferroelectrics\cite{ViznerStern2021,Xue2021} provide versatile building blocks to design valves that allow controlling the electronic properties of van der Waals heterostructures.\cite{2022arXiv220504458K,Marrazzo2022,Si2019,Chen2021}

Graphene multilayers 
provide a flexible family of heterostructures
to realize correlated, superconducting, and topological states.\cite{Cao2018_CI,Cao2018_SC,doi:10.1126/science.aaw3780,doi:10.1126/science.aay5533,Choi2021} 
In the particular case of AB-stacking bilayer graphene, an electric bias between the layers opens an energy gap, leading to an electrically-tunable gap semiconductor,\cite{PhysRevLett.99.216802,Oostinga2008,Zhang2009,Xia2010,PhysRevX.8.031023} and to a superconducting state in the slightly doped regime\cite{Zhou2022}. 
At half-filling,
the valleys at K and K' carry a momentum-space Berry curvature with opposite sign due to the broken inversion symmetry that occurs in this gapped system, thus displaying a valley Hall insulator behavior.\cite{PhysRevLett.100.036804,Ju2015,Zhang2013,PhysRevX.3.021018} This phenomenon has been exploited to create topological edge modes in the junction between oppositely gated regions\cite{Qiao2011} and also in AB-BA stacking domain walls.\cite{Ju2015} In the absence of valley mixing, these chiral modes are topologically protected against back-scattering and present a ballistic transport.\cite{Li2016,Rickhaus2018}  
While such phenomena have been induced by an external  bias, ferroelectric-based van der Waals heterostructures would lead to intrinsically persistent topological channels driven by the ferroelectric order. The control on the ferroelectric domains of the Valley Hall valve realized in such heterostructure would allow generating and tuning topological excitations. Therefore, it would provide a promising platform to print circuits displaying ballistic behavior.

\begin{figure*}[t!]
  \centering
  \includegraphics[width=\textwidth]{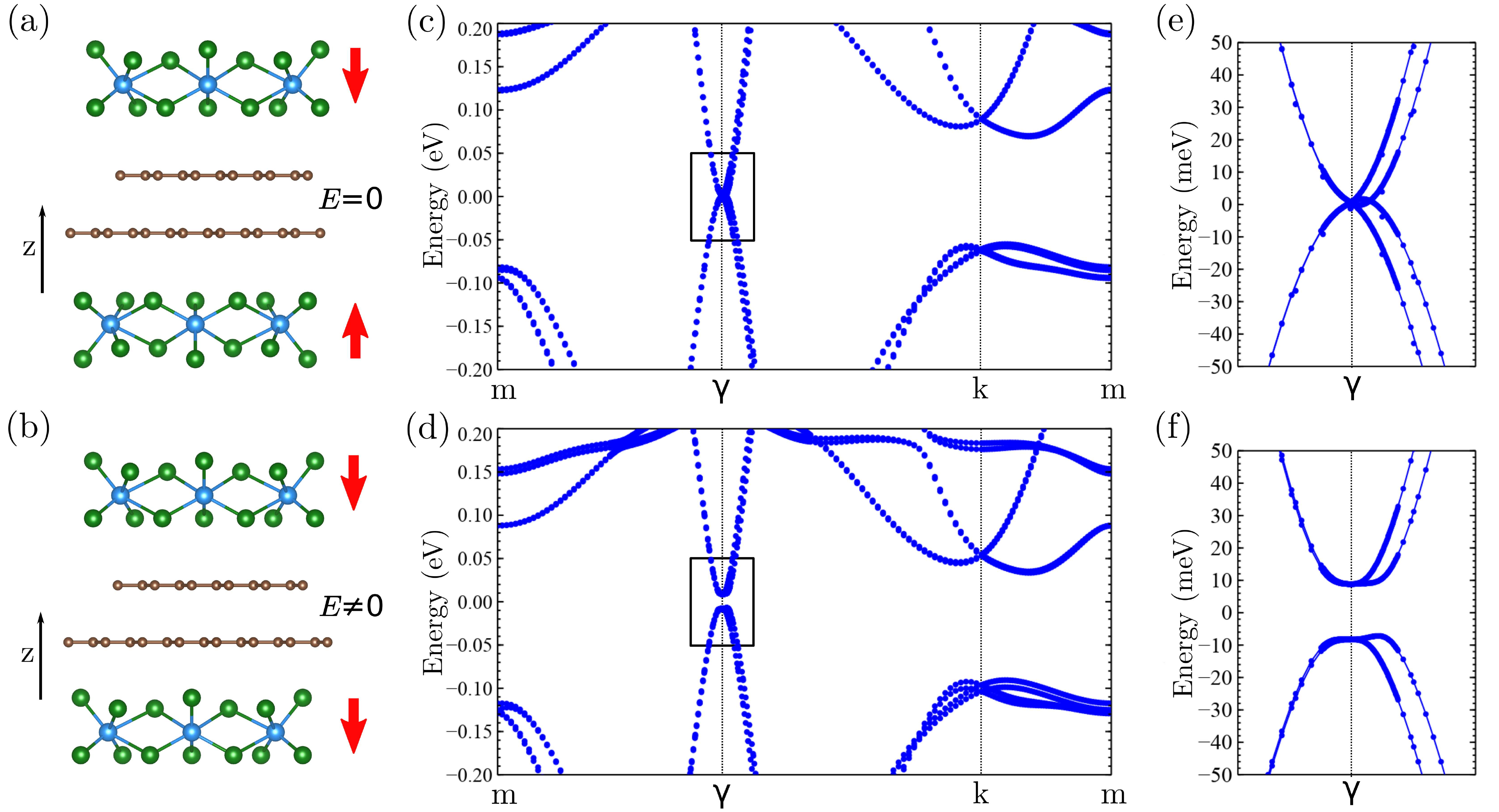}
     \caption{(a,b) Structures of AB bilayer graphene encapsulated in ferroelectric MoTe$_2$ used in DFT calculations. In configuration (a), the ferroelectric order in MoTe$_2$ leads to a net zero electric field in the graphene bilayer, while in configuration (b) the ferroelectric proximity of the encapsulation leads to a non-zero electric field. The red arrows show the direction of the ferroelectric polarization in each of the MoTe$_2$ layers. C, Mo, and Te atoms are depicted in brown, blue, and green respectively. (c,d) Band structure calculations for the structures that are shown in panels (a) and (b) respectively. In panel (c) no energy gap opens as the net electric field is zero, while in panel (d) an energy gap opens up due to the net electric field induced in the graphene bilayer. Panels (e) and (f) show the zoomed areas near the Fermi level depicted as rectangles in panels (c) and (d) respectively.}\label{Fig:dft_bands}
\end{figure*}

In this work, we demonstrate, using a combination of \emph{ab initio} calculations and low-energy models, the emergence
of a valley valve in MoTe$_2$/bilayer graphene/MoTe$_2$ van der Waals heterostructures. The ferroelectric order of MoTe$_2$ gives rise to a built-in electrostatic bias in bilayer graphene,
controllable by the orientation between the two MoTe$_2$ layers. In particular, the relative orientation
of the ferroelectric order in MoTe$_2$ allows switching on and off the electronic gap of bilayer graphene and controlling its topological gap.
In the insulating regime, topological states emerge in the boundary between opposite ferroelectric domains. We analyze the effect of the moiré modulation in the interlayer potential driven by the mismatch and potential twist between graphene and MoTe$_2$. Our results show that MoTe$_2$/graphene heterostructures realize robust valley topological modes in the presence of moiré modulations, thus putting forward a promising platform to obtain a ferroelectric valley valve with electrically-controlled topological excitations.

\section{Results and Discussion}

We start demonstrating using \emph{ab initio} first principles methods that bilayer graphene encapsulated in MoTe$_2$ displays a controlled gap driven by the ferroelectric order. Then, we will analyze with a low energy model the emergence of topological excitations in the boundary between two ferroelectric domains and their robustness against the moiré modulations induced by the mismatch between MoTe$_2$ and graphene.

\subsection{Ab initio calculations}

MoTe$_2$ shows a ferroelectric order with an out-of-plane electric polarization in the distorted d1T phase, which has been proven to be stable in MoTe$_2$/graphene heterostructures.\cite{Yuan2019}  We will exploit these properties to create a van der Waals heterostructure in which an electrically-controlled valley valve emerges. Our heterostructure consist
of monolayer MoTe$_2$, Bernal-stacked AB graphene, and monolayer MoTe$_2$ as shown in Fig. \ref{Fig:dft_bands}ab.
Figures \ref{Fig:dft_bands}a and \ref{Fig:dft_bands}b show the two possible configurations
of the ferroelectric heterostructure that determine the valve. In the first configuration, the electric polarization of each of the MoTe$_2$ layers can be oriented in opposite direction leading to a zero electric field $E=0$ in the heterostructure (Fig. \ref{Fig:dft_bands}a).
In the second configuration, the electric polarizations of the MoTe$_2$ layers point in the same direction producing a non-zero electric field $E\neq 0$ in the heterostructure that can be felt by the graphene bilayer (Fig. \ref{Fig:dft_bands}b).

We now analyze the electronic structure of both configurations using first-principles 
density functional theory (DFT).\cite{HK}
To create a computationally affordable heterostructure, we have commensurated this lattice parameter $a$ with a 3$\times$3 supercell for graphene and fix $a=7.38$ Å.\footnote{An alternative ratio between the lattice parameters of pristine graphene and MoTe$_2$ would be 5$\times$5 supercells of graphene embedded in 2$\times$2 supercells of MoTe$_2$. We analyze in the next section the effect that such 5$\times$5 modulation creates in the electronic properties of the heterostructure.} 
DFT calculations are performed with {\sc Quantum ESPRESSO}.\cite{0953-8984-21-39-395502,0953-8984-29-46-465901} 
These calculations were converged concerning all parameters in a 12$\times$12 $k$-mesh and introducing a vacuum of 20 Å in the z-direction to ensure that there is not interaction of the heterostructure with neighboring unit cells in the z-direction.

\begin{figure}[t!]
  \centering
  \includegraphics[width=\columnwidth]
        {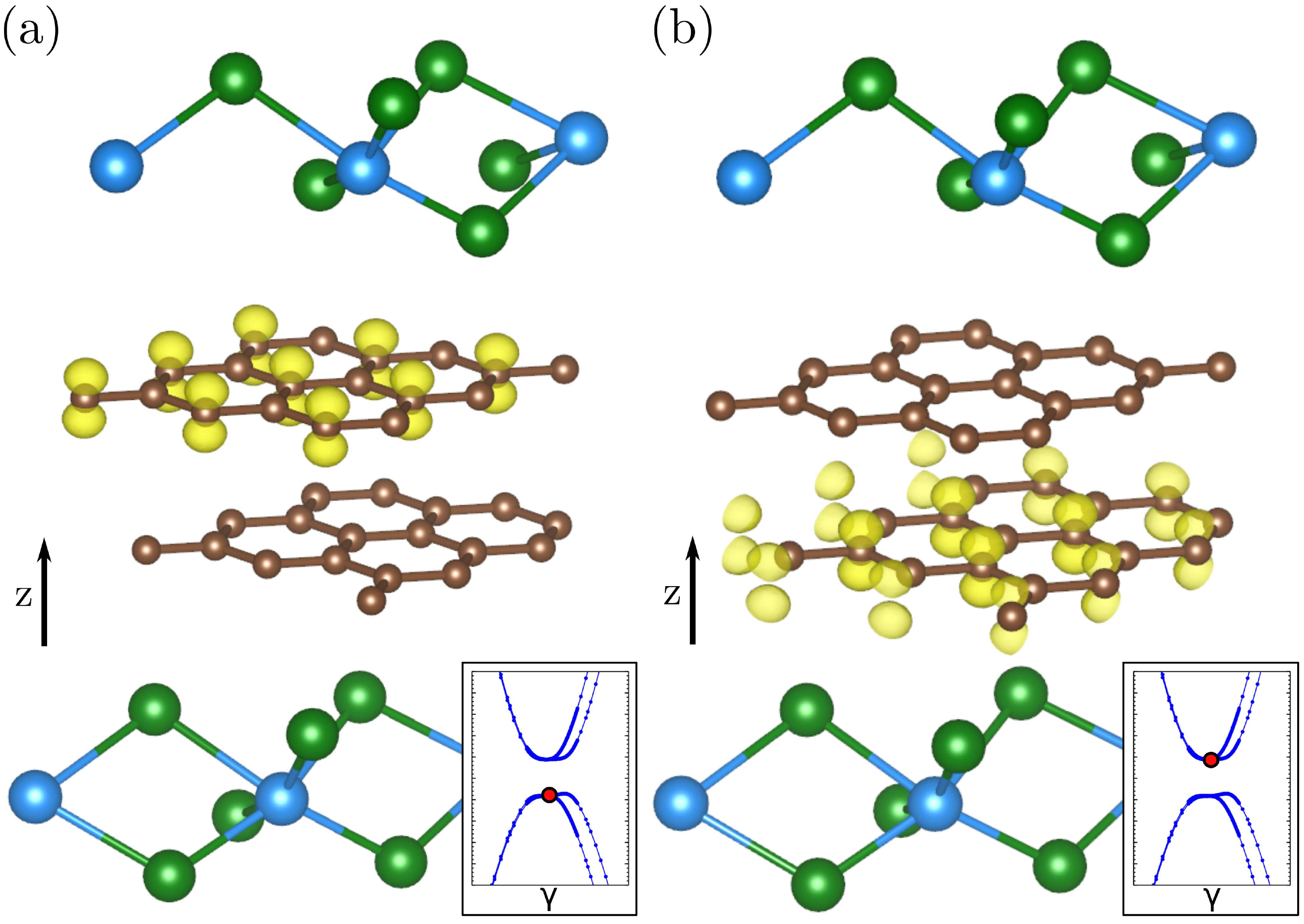}
     \caption{Wavefunction probability density (yellow isosurface) of the highest occupied orbital (a) and the lowest unoccupied orbital (b) for the gapped configuration (Fig. \ref{Fig:dft_bands}b) at the $\gamma$ point, demonstrating the valley Hall insulating behavior. C, Mo, and Te atoms are depicted in brown, blue, and green respectively. The insets show as red dots in the band structure the chosen wavefunctions to be plotted. Wavefunctions of the highest occupied orbital (a) (lowest unoccupied orbital (b)) are localized in the C atoms of the top (bottom) layer of graphene, in the carbon atoms that do not sit on top of another carbon atom in the z-direction.}
     \label{Fig:dft_wf}
\end{figure}

Figures \ref{Fig:dft_bands}c and \ref{Fig:dft_bands}d show the corresponding band structures for both configurations of the heterostructure (Figs. \ref{Fig:dft_bands}a and \ref{Fig:dft_bands}b respectively).
We can see in Fig. \ref{Fig:dft_bands}c, corresponding to the configuration with $E=0$, that the
electronic structure of the graphene bilayer remains gapless. This can be better seen in Fig. \ref{Fig:dft_bands}e which corresponds to a zoom of the rectangular area in Fig. \ref{Fig:dft_bands}c. Nevertheless, the band structure for the configuration with $E\neq0$ displays a different behavior. We can see in Fig. \ref{Fig:dft_bands}d that an energy gap opens at the Fermi level as a consequence of the
ferroelectric-induced field in the z-direction felt by the graphene bilayers. In the zoomed band structure (Fig. \ref{Fig:dft_bands}f) we can observe more clearly the induced gap, which takes a value $\Delta_E \approx 15$ meV. We note that in this structure no external field is applied, and the opening stems
solely from the proximity effect of the ferroelectric.
From the DFT calculations, we have also obtained that the size of the electric dipole in the gapped heterostructure is $p=0.4$ Debye, which leads to an electric bias $\Delta V=22$ mV between the graphene layers. Note that this bias is two or three orders of magnitude smaller than the one externally applied in AB bilayer graphene devices to obtain the same insulating behavior.\cite{PhysRevLett.99.216802,Oostinga2008} In those devices a large screening of the electric voltage occurs in comparison to the bilayer graphene encapsulated in MoTe$_2$, where the ferroelectric directly gates the bilayer without screening.

To demonstrate the topological nature of the gap in the insulating configuration, we have computed the wavefunction probability density of the highest occupied and the lowest unoccupied orbitals (see Figs. \ref{Fig:dft_wf}a and \ref{Fig:dft_wf}b and the insets to identify the corresponding wavefunctions). The wavefunctions of the highest occupied orbital are localized in the carbon atoms of the top graphene layer that do not lie on top of the carbon atoms of the other layer in the AB stacking (Fig. \ref{Fig:dft_wf}a). In the case of the lowest unoccupied orbital the wavefunctions are localized in the carbon atoms of the bottom graphene layer that, again, are not aligned with the carbon atoms of the top layer in the AB stacking (Fig. \ref{Fig:dft_wf}b). This feature demonstrates that
the gap opening stems from an electrostatic imbalance between the two graphene layers, and not from intervalley scattering
in the $3\times3$ supercell,
equivalent to the effect of an externally applied bias.\cite{PhysRevLett.100.036804}

\subsection{Low-energy model}

We now analyze the emergence of topological excitations in the domain walls that can be created in this ferroelectric heterostructure. To do so, we will effectively describe the bilayer graphene encapsulated in MoTe$_2$ with a low-energy model. 
The electronic structure of the heterostructure around the Fermi level can be described with a tight-binding Hamiltonian for AB bilayer graphene, where the effect of the ferroelectric is incorporated
through its layer-dependent electrostatic potential.

\begin{equation}\label{eq:hamiltonian}
 H= t\sum_{\langle ij\rangle }c_{i}^{\dagger}c_{j}+ t_{I}\sum_{ ij }\nu_{ij} c_{i}^{\dagger}c_{j}+ E\sum_{i}\beta_i c_{i}^{\dagger}c_{i}, 
\end{equation}

where $t$ is the first neighbor intralayer hopping, $t_I$ is the interlayer hopping, and
$i$ and $j$ run over the atomic positions.
The term $\nu_{ij}$ takes
value $\nu_{ij}=1$ for $i,j$ belonging to different
layers and stacked right on top of each other, and $\nu_{ij}=0$ otherwise. The term $\beta_i$ takes value $\beta_i=1$
for the top layer and $\beta_i=-1$ for the bottom layer.
Finally, $E$ is the electrostatic difference
between the two layers, which depends on the relative configuration of the ferroelectrics.
As a reference, the experimental value of the hoppings for graphene bilayers are $t_I \approx 0.15t$,
corresponding to $t\approx 3$ eV and $t_I \approx 450$ meV.\cite{RevModPhys.81.109}

\begin{figure}[t!]
  \centering
  \includegraphics[width=\columnwidth]
        {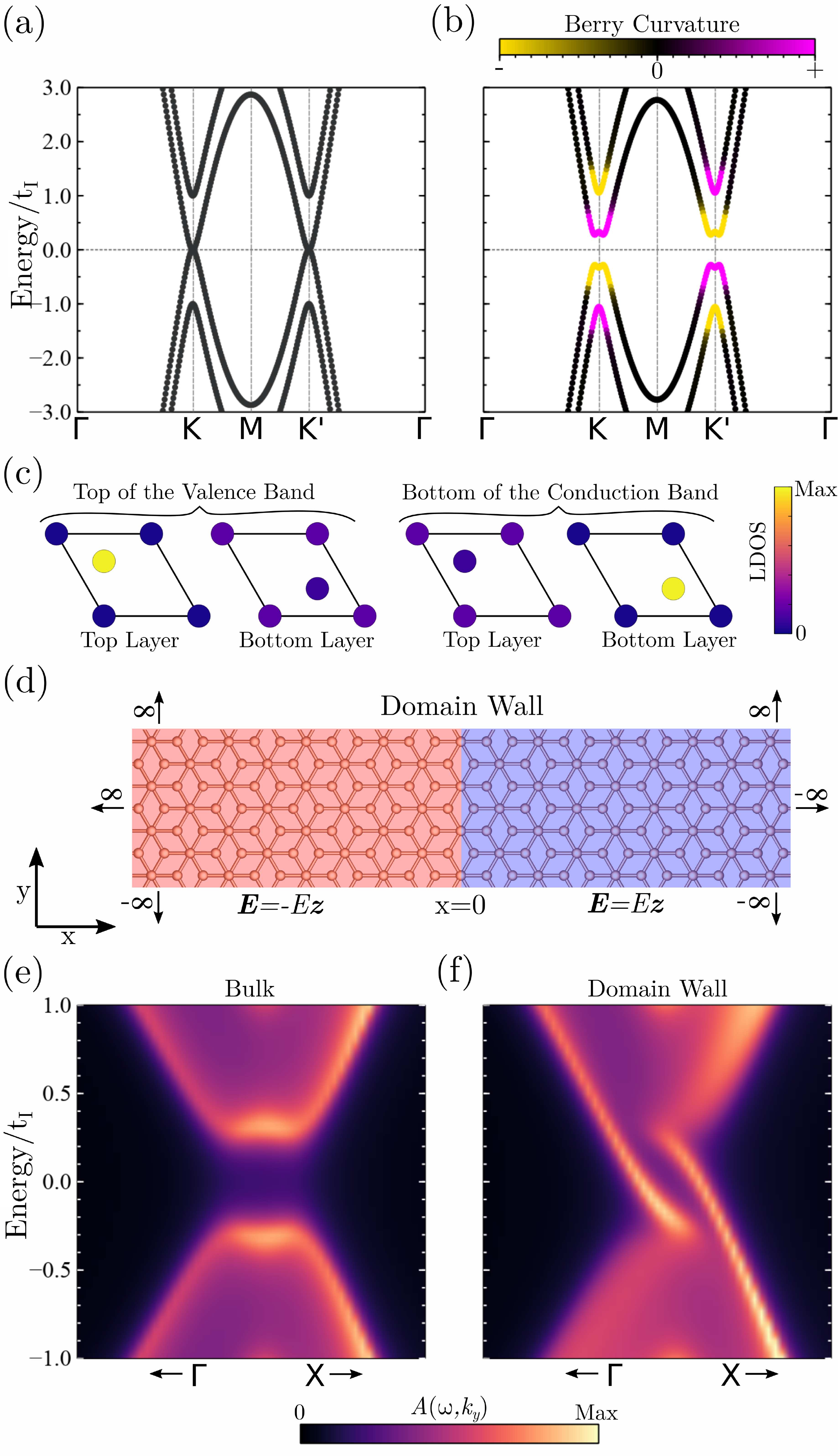}
     \caption{Low energy model electronic structure of AB bilayer graphene under the absence/presence of an electrostatic bias between layers (eq. (\ref{eq:hamiltonian})) (a) Band structure int he absence
     of interlayer bias ($E/t=0$), corresponding to the heterostructure configuration of Fig. \ref{Fig:dft_bands}a.  (b) Band structure in the presence of a finite interlayer bias ($E/t=0.1$), 
     corresponding to the heterostructure configuration of Fig. \ref{Fig:dft_bands}b. The Berry curvature has also been plotted in the bands. (c) Local density of states (LDOS) for  $E/t=0.1$ at the top of the valence band and at the bottom of the conduction band. (d) Schematic of a boundary between different ferroelectric domains (highlighted in red and blue). Both domains display the valley Hall insulating configuration ( Fig. \ref{Fig:dft_bands}b), but the electric fields of each domain point in the opposite direction. The domains are semi-infinite in the $x$ direction. The domain wall occurs at $x=0$. In the $y$ direction the lattice is periodic. 
     (e,f) Momentum resolved spectral function $A(\omega, k_y)$ for the bulk domains (e) and for the domain wall (f).}
     \label{Fig:low_E_model}
\end{figure}

Figure \ref{Fig:low_E_model}a shows the band structure for the low energy model (eq. (\ref{eq:hamiltonian})) for $E/t=0$, this situation corresponds to the heterostructure configuration shown in Fig. \ref{Fig:dft_bands}a, where the net internal interlayer bias induced by the MoTe$_2$ is zero.  
We can observe that in the absence of an electric field there is no energy gap, in agreement with the DFT band structure (Fig. \ref{Fig:dft_bands}e). Figure \ref{Fig:low_E_model}b shows the low energy-model band structure for  $E/t=0.1$, this situation corresponds to the heterostructure configuration shown Fig. \ref{Fig:dft_bands}b. In this case, the induced internal
interlayer bias in bilayer graphene is non-zero and a gap opens as in Fig. \ref{Fig:dft_bands}f. 
The local density of states for $E/t=0.1$ is plotted in Fig. \ref{Fig:low_E_model}c. We can see that the top of the valence band LDOS localizes in the top-layer C atom that does not align with the carbon from the bottom layer. For the bottom of the conduction band LDOS, it localizes in the bottom-layer C atom that does not align with the carbon of the top layer. This result is in agreement with the DFT wavefunction probability density plots shown in Fig. \ref{Fig:dft_wf}, thus confirming the valley Hall insulating character of bilayer graphene encapsulated in MoTe$_2$.

Now that we have demonstrated that the low energy model provides a good description of the DFT calculations that we have performed in the heterostructure, we can analyze with it the topological excitations that can emerge.
The Berry curvature associated with the energy bands for the insulating case is shown in Fig. \ref{Fig:low_E_model}b. It is observed that the Berry curvature is localized in the $K$ and $K'$ points,
which allows defining a valley Chern number $\mathcal{C}_K$ and $\mathcal{C}_{K'}$. 
In the biased bilayer, each valley leads to a Chern number $\mathcal{C}_{K,K'} = \pm 1$,
leading to a net valley Chern number
$\mathcal{C}_{V} = \mathcal{C}_{K} - \mathcal{C}_{K'}=\pm 2$.
The existence of a finite valley Chern number
gives rise to valley polarized gapless edge modes at the interface between
two insulating domains with opposite valley Chern numbers. Figure \ref{Fig:low_E_model}d shows a schematic of a domain wall between two domains (red and blue) in which the electric field $E$ felt by the bilayer has the opposite direction. This corresponds to a situation in which the electric polarization of MoTe$_2$ layers induces a finite electric field in bilayer graphene, but in each of the domains this field takes the opposite direction. 
Therefore, the valley Chern numbers take an opposite sign in each of the domains at $K$ and $K'$. The emergence of interface states is determined by the difference between the corresponding valley Chern numbers at  $K$ and $K'$ between red and blue domains. Figures \ref{Fig:low_E_model}e and \ref{Fig:low_E_model}f show the momentum resolved spectral function $A(\omega, k_y)$ for the domain system described in Fig. \ref{Fig:low_E_model}d. We can see that the bulk of both domains remains gapped (Fig. \ref{Fig:low_E_model}e), but at the domain wall (\ref{Fig:low_E_model}f) two zero-energy modes emerge as a consequence of the opposite sign that the valley Chern numbers display in each domain.\cite{PhysRevLett.100.036804} 
Interestingly, externally controlling ferroelectric domains allows directly imprinting topological excitations in the heterostructure. Ultimately, the control of the ferroelectric domains allows to use this feature to imprint topological circuits with a ballistic transport behavior in graphene/MoTe$_2$ heterostructures without requiring to externally apply a bias.

\begin{figure*}[t!]
  \centering
  \includegraphics[width=0.9\textwidth]
        {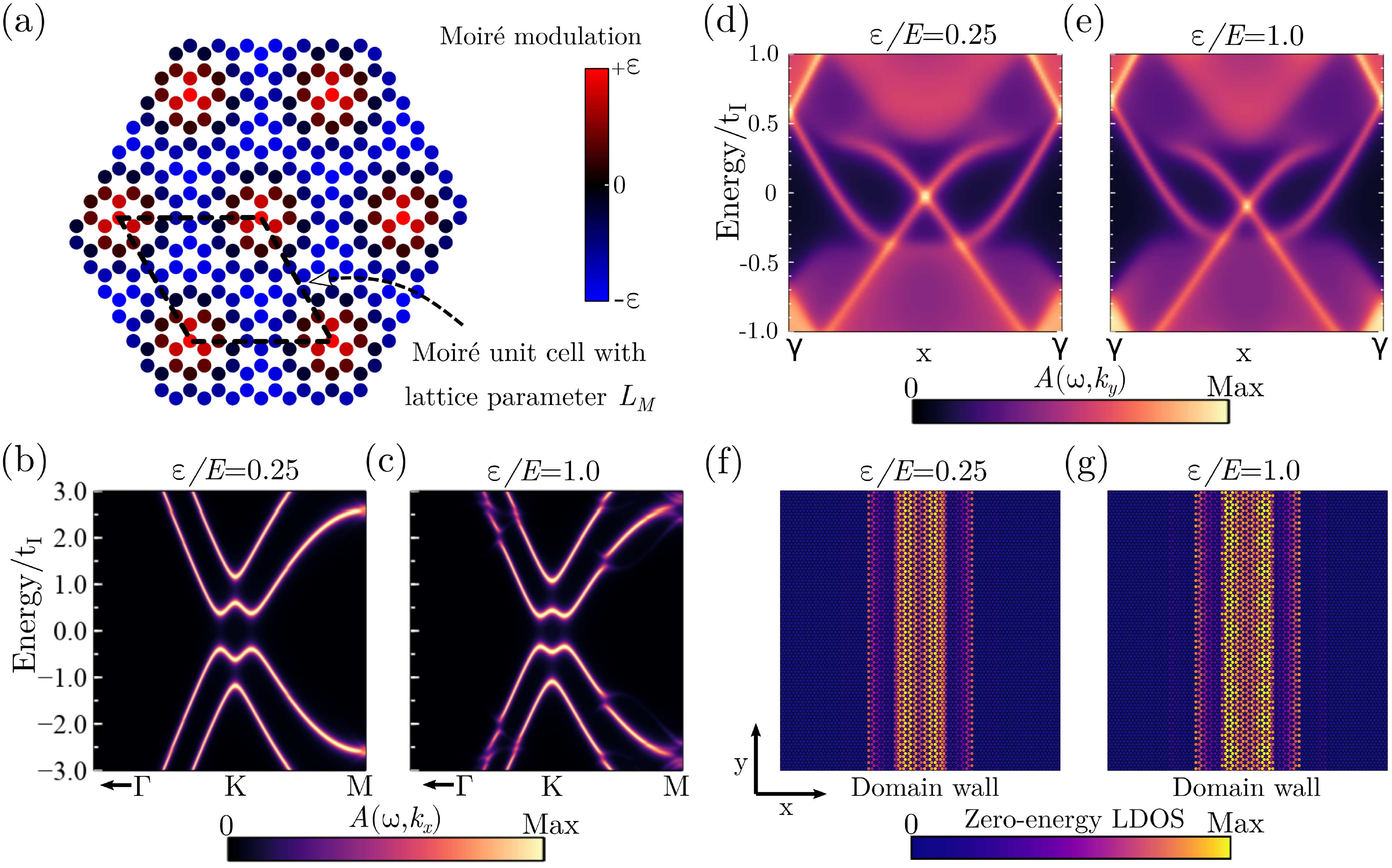}
     \caption{Calculations including a moiré potential (eq. (\ref{eq:moire_potential})) in the low energy model of eq. (\ref{eq:hamiltonian}) for $E/t=0.1$. (a) Induced moiré potential by the MoTe$_2$ in each of the graphene layers. A moiré unit cell emerges on top of the original bilayer graphene unit cell. (b,c) Unfolded momentum resolved spectral functions for $\epsilon/E=0.25$ (b) and $\epsilon/E=1.0$ (c). (d,e,f,g) Calculations at the interface between the two ferroelectric domains shown in Fig. \ref{Fig:low_E_model}d, they include the effect of the moiré potential. (d,e) Momentum resolved spectral function for: (d) $\epsilon/E=0.25$ and (e) and $\epsilon/E=1.0$. (f,g) Zero-energy local density of states (LDOS) for: (f) $\epsilon/E=0.25$ and (g) and $\epsilon/E=1.0$. }
     \label{Fig:moire}
\end{figure*}

Finally, we discuss the impact of the moiré modulation that emerges due to the mismatch and rotation
between bilayer graphene and MoTe$_2$. Such a moiré modulation arises naturally as a consequence of the lattice-parameter mismatch between MoTe$_2$ and graphene, and it can be controlled by adjusting the twist angle between the layers of both van der Waals materials.
The effect of the moiré modulation can be directly accounted for in the low energy model (eq. (\ref{eq:hamiltonian})),
by including a spatially dependent electrostatic potential
of the form

\begin{equation}\label{eq:moire_potential}
   H_{M}=
    \sum_{i}
    \epsilon(\mathbf r_i)
    \beta_i c_{i}^{\dagger}c_{i},
\end{equation}

where $\epsilon(\mathbf r_i) = \epsilon\sum_{\alpha} \cos\left(\frac{\mathbf{b}_\alpha\cdot\mathbf{r}_i}{n}\right) +C_0 $, with
$\mathbf{b}_\alpha$ the reciprocal lattice vectors of the moiré unit cell (the summation runs over the 3 $\mathbf{b}_\alpha$ vectors related by C$_3$ symmetry).
The parameter $C_0$ is taken so that $\langle \epsilon(\mathbf r_i) \rangle = 0$.
The product $\mathbf{b}_\alpha\cdot\mathbf{r}_i$ equals $2\pi$ when $\mathbf{r}$ takes the value of the bilayer graphene lattice vectors, and $n$ is an integer that commensurates the moiré length $L_{M}$ with the original lattice parameter of graphene $a_C$ as $L_{M}=na_C$. Therefore, Eq. (\ref{eq:moire_potential}) allows to include a modulated electrostatic
potential as the one shown in Fig. \ref{Fig:moire}a in each of the graphene layers. For concreteness,
we will analyze the case $L_{M}=5a_C$, which corresponds
to 
a 5$\times$5 supercell for
graphene,\footnote{We have checked that the phenomenology presented here is robust against other modulations different from the 5$\times$5.} with a well 
approximated commensurability (less than 1\% difference) with MoTe$_2$. 
The strength of the modulation is given by $\epsilon$. We will analyze a moderate
modulation of the electric bias $\epsilon/E=0.25$, and the limit case where
the modulation amplitude equals the electric bias $\epsilon/E=1.0$.

Figures \ref{Fig:moire}b and \ref{Fig:moire}c show the unfolded momentum resolved spectral functions for $\epsilon/E=0.25$ (\ref{Fig:moire}b) and $\epsilon/E=1.0$ (\ref{Fig:moire}c) in the original bilayer graphene unit cell. We can observe the emergence of band anticrossings most noticeable for the strong modulated case ($\epsilon/E=1.0$). However, the energy gap remains open and barely affected by the moiré modulation. Therefore, we might expect that the valley Hall insulating behavior is robust against possible moiré modulations that might appear between graphene and MoTe$_2$ layers. To confirm this, Figs. \ref{Fig:moire}d and \ref{Fig:moire}e show the momentum resolved spectral function at the interface between the two ferroelectric domains described in Fig. \ref{Fig:low_E_model}d, this time including the effect of the moiré potential for $\epsilon/E=0.25$ (Fig. \ref{Fig:moire}d) and $\epsilon/E=1.0$ (Fig. \ref{Fig:moire}e). Note that in this case the bands have not been unfolded. We can see that zero-energy modes emerge at the interface of the domain walls even for the strongly modulated potential ($\epsilon/E=1.0$).
The effect of the moiré potential in the zero-energy modes can be further rationalized in real space, as shown in Figs. \ref{Fig:moire}f and \ref{Fig:moire}g, 
where the zero-energy local density of states (LDOS) is shown at the interface between the two ferroelectric domains described in Fig. \ref{Fig:low_E_model}d.  First, it is observed that the zero-energy modes localize at the domain wall, as expected
from their in-gap nature. Furthermore, for a strong moiré modulation,
($\epsilon/E=1.0$, Fig. \ref{Fig:moire}g) the interface modes
show a modulation in real space following the moiré potential.\cite{Kezilebieke2022,PhysRevMaterials.6.094010}
It is worth noting that such a moire modulation does not lead to intervalley scattering, as demonstrated
by the gapless nature of the spectral function of Fig. \ref{Fig:moire}d.
Therefore, these results demonstrate that the valley Hall insulating
character displayed by the graphene-MoTe$_2$ heterostructure is robust
against moiré modulations between the ferroelectric and graphene layers, meaning that
a finite misalignment between the layers can exist yielding to the same phenomenology, a feature
that dramatically simplifies the experimental design of these heterostructures.

\section{Conclusions}

To summarize, we have demonstrated an emergent ferroelectrically-driven valley Hall insulating behavior in AB-stacking bilayer graphene encapsulated in ferroelectric MoTe$_2$ using a combination of \emph{ab initio} calculations and low-energy models. First-principles calculations show how the gap in bilayer graphene can be opened and closed as a function of the orientation between the two layers of MoTe$_2$. For opposite ferroelectric polarization between the MoTe$_2$ encapsulation, the net effective field felt by bilayer graphene is zero and hence the system remains gapless. When the electric polarizations of both MoTe$_2$ layers point in the same direction, a non-zero 
electric field is induced in graphene bilayer, 
thus producing a valley Hall insulating behavior. Furthermore, we demonstrated
that moiré modulations of the electrostatic bias created by the ferroelectric do not impact
the valley Hall insulating character of the heterostructure, and topological modes are robust to it. 
Our results put forward ferroelectric/graphene
heterostructures as versatile valves with a
built-in electrostatically controlled topological
electronic structure. 
Ultimately,
the built-in electric fields of these heterostructures
would allow creating a wide variety of permanent
electrically defined quantum dots architectures
and biased twisted graphene multilayers, without
requiring the application of external biases.

\section*{Acknowledgements}
We acknowledge the computational resources provided by
the Aalto Science-IT project,
and the financial support from the
Academy of Finland Projects No. 331342, No. 336243 and No. 349696
and the Jane and Aatos Erkko Foundation. 
We thank. P. Liljeroth, M. Amini and S. Kezilebieke
for useful discussions.

\end{document}